\documentclass[prl,twocolumn,showpacs,preprintnumbers,amsmath,amssymb,footinbib]{revtex4}% PRL

\usepackage{graphicx}% Include figure files
\usepackage{dcolumn}% Align table columns on decimal point
\usepackage{bm}% bold math

\begin{document}

%\preprint{APS/123-QED}

\title{The Isotropic to Nematic Liquid Crystalline Phase Transition of F-actin Varies from Continuous to First Order} % Force line breaks with \\

\author{Jorge Viamontes}
\author{Patrick W. Oakes}
\author{Jay X. Tang} 
\email[Correspondence address:]{Jay_Tang@Brown.edu}
\affiliation{Department of Physics, Brown University, Rhode Island 02912, USA}

\date{\today}	% It is always \today, today,
             %  but any date may be explicitly specified

\begin{abstract}
We report that the properties of the isotropic to nematic liquid crystalline phase transition of F-actin depend critically on the average filament length.  For average filament lengths longer than 2 $\mu$m, we confirm previous findings that the phase transition is continuous in both alignment and concentration.  For average filament lengths shorter than 2 $\mu$m, we show for the first time a first order transition with a clear discontinuity in both alignment and concentration.  Tactoidal droplets of coexisting isotropic and nematic phases, differing in concentration by approximately 30\%, form over the course of hours and appear to settle into near equilibrium metastable states.  
\end{abstract}

\pacs{61.30.Eb, 64.70.Md, 87.15.-v}% PACS, the Physics and Astronomy
                             % Classification Scheme.
\maketitle

The cytoskeletal protein actin is vital to both cell morphology and motility.  In its globular form, G-actin, the protein can be polymerized to form long filaments, F-actin. These filaments have a diameter of 8 nm \cite{HolmesPoppGebhardKabsch} and a distribution of lengths characteristic of the stochastic polymerization process \cite{Sept_Xu_Pollard_McCammon}. They behave as semiflexible polymers with persistence lengths of 15-18 $\mu$m \cite{Isambert_Vernier_Maggs_Carlier,Gittes_Mickey_Nettlenton_Howard}, which are larger than the average lengths found in cells and \textit{in vitro}. Due to the important role actin plays in biological functions there have been extensive studies concerning many properties of F-actin, including dynamic filament assembly and dissembly \cite{Fujiwara2002}, phase transitions \cite{Coppin, Furakawa, Viamontes2003}, and viscoelasticity \cite{GardelPRL2003,XuPalmerWirtz,IsambertMaggs}. 

Of particular relevance to this report is that F-actin undergoes an isotropic (I) to nematic (N) liquid crystalline phase transition. Understanding this thermodynamic property in the biological process is essential because of the role that filament alignment plays in cellular dynamics \cite{Sept_Xu_Pollard_McCammon} and possibly the anisotropic viscoelasticity of the cytoskeletal network \cite{XuPalmerWirtz}.  The onset concentration of the I-N transition of F-actin is inversely proportional to the average filament length $\ell$ \cite{Coppin, Furakawa, Suzuki, Viamontes2003}, consistent with statistical mechanical theories \cite{Onsager,Flory}. Experimental studies use optical birefringence methods to measure the F-actin alignment across the I-N transition region. Under certain preparation conditions, zebra birefringence patterns were observed, which have been attributed to the spontaneous separation of F-actin into I and N domains \cite{Suzuki}.  More recently, however, two studies \cite{Coppin, Viamontes2003} have shown that the F-actin I-N transition appears to be continuous in both filament alignment and concentration for $\ell \geq$ 2 $\mu$m. The large length, polydispersity and semi-flexibility of the filaments may contribute to entangling the network, ultimately resulting in a continuous transition \cite{Viamontes2003}. This phenomenon may also be relevant to the theory of Lammert, Rokshar, and Toner (LRT) \cite{LRTa,LRTb}, which predicts that a high disclination line defect energy renders the I-N transition into two continuous ones.

In this paper we confirm the continuous features of the I-N transition for solutions of long F-actin ($\ell \geq$ 2 $\mu$m), but also show for the first time a true first order phase separation for solutions of F-actin with $\ell \leq$ 2 $\mu$m. Up to this point a clear separation of phases into distinct domains has not been observed in solutions of F-actin.  Tactoidal droplets characteristic of phase separation in such a rod-like system were observed in at least three types: N tactoids in an I background, I tactoids in an N background, and the coexistence of N and I tactoids in a background of weak alignment. Quantitative fluorescence measurements show a distinct difference in concentration between the I and N regions.  In addition birefringence measurements show a difference in alignment, further demonstrating the first order nature of the transition.

\begin{figure*}[t]
\includegraphics[width=17.5cm]{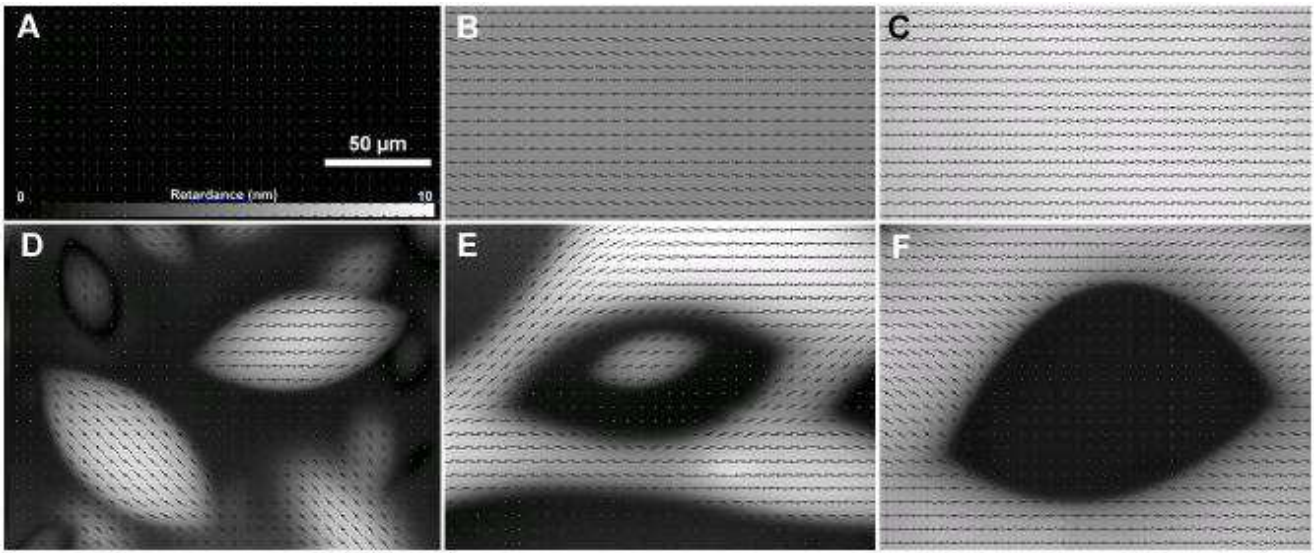}
\caption{\label{fig1}Birefringence measurement of F-actin solutions at $\ell$ = 11 $\mu$m (\textbf{A,B,C}) and at $\ell$ = 1 $\mu$m (\textbf{D,E,F}). The 50 $\mu$m scale bar and the gray scale intensity bar, indicating retardance values from 0 to 10 nm, apply to all the images. The line segments represent the local direction and relative magnitude of filament alignment. (\textbf{A},\textbf{B},\textbf{C}) Representative samples are shown in the I phase, in the I-N transition region for long F-actin, and in the N phase, respectively. No discontinuity is detected in the transition region. (\textbf{D},\textbf{E},\textbf{F}) At $\ell$ = 1 $\mu$m and in the I-N transition region, F-actin phase separates into tactoidal droplets: N droplet in an I background (\textbf{D}), coexistence of N and I droplets (\textbf{E}), and an I droplet in a N background (\textbf{F}).}
\end{figure*}

Actin was extracted from rabbit skeletal muscle following an established method \cite{Pardee}.  The globular G-actin was polymerized upon addition of KCl and MgCl$_2$ up to 50 mM and 2 mM, respectively.  The $\ell$ of F-actin was controlled through the addition of gelsolin, a filament severing and end-capping protein \cite{JanmeyPeetermans,TangJanmey}.  Rectangular capillary tubes from VitroCom Inc. (Mt. Lks., N.J) of cross-sectional dimensions 0.2$\times$2 mm were used for measurements by fluorescence and birefringence microscopy. Both ends of the capillary tube were sealed by an inert glue to eliminate flow. Birefringence measurements were performed on a Nikon E-800 microscope equipped with the CRI PolScope (Cambridge, MA) package \cite{Viamontes2003}. PolScope is capable of measuring the optical birefringence and the direction of the slow axis at each pixel position, thus reporting local alignment \cite{OldenbourgMei}.  For fluorescence imaging F-actin was labeled 1 to 1000 with TRITC-phalloidin (Sigma, St Louis, MO) as previously described \cite{Viamontes2003}.  

Different features are observed between samples of several $\ell$ of F-actin in their respective ranges of concentration over which the I-N transition occurs. A sample of F-actin with no gelsolin added was measured to have an $\ell$ = 11 $\mu$m, a value representative of most of our samples. Fig.~\ref{fig1} shows typical results of birefringence and filament alignment for F-actin in the I phase (Fig.~\ref{fig1}\textbf{A}), transition region (\ref{fig1}\textbf{B}), and the N phase (\ref{fig1}\textbf{C}). Of particular note is that uniform retardance is found in the I-N transition region (\ref{fig1}\textbf{B}).  This suggests that F-actin is continuous in alignment during the transition in addition to being continuous in concentration as found previously \cite{Viamontes2003}. Zebra patterns were occasionally observed, especially near the wall of a capillary tube or at an air liquid interface, examples of which have been shown by previous studies \cite{Coppin, Suzuki, Viamontes2003,Helfer_Panine_Carlier}. These zebra patterns, representing regions of different alignment, do not show a difference in concentration.  Coexistence is never observed in long F-actin samples \cite{Viamontes2003, Das, Helfer_Panine_Carlier}. In stark contrast, however, further reduction of $\ell$ to $\leq$ 2 $\mu$m results in the F-actin solution phase separating into tactoidal droplets and a surrounding medium (Fig.~\ref{fig1}\textbf{D}). A small increase in either concentration or average filament length can then give rise to co-existence of I and N tactoids (\ref{fig1}\textbf{E}), and I tactoids in a N background (\ref{fig1}\textbf{F}). 

While the formation of tactoids is a robust process, the tactoids themselves are sensitive to mechanical perturbations.  Even gentle shaking distorts the boundary between phases.  Strong shaking or sonication almost entirely distorts the order in the system.  When the agitation is stopped the system again phase separates and settles back into a stable state over a period of a few hours.  Tactoids are almost always present in this resettled system.  Once stable, the system remains so for days.  It is important to note that, while being sensitive to mechanical perturbations, this phase separation does not necessarily require the initial alignment created by injection of the filaments into the capillary.  In fact, tactoids have also been observed on slides where there is no initial alignment.  

\begin{figure}[t]
\includegraphics[width=8.5cm]{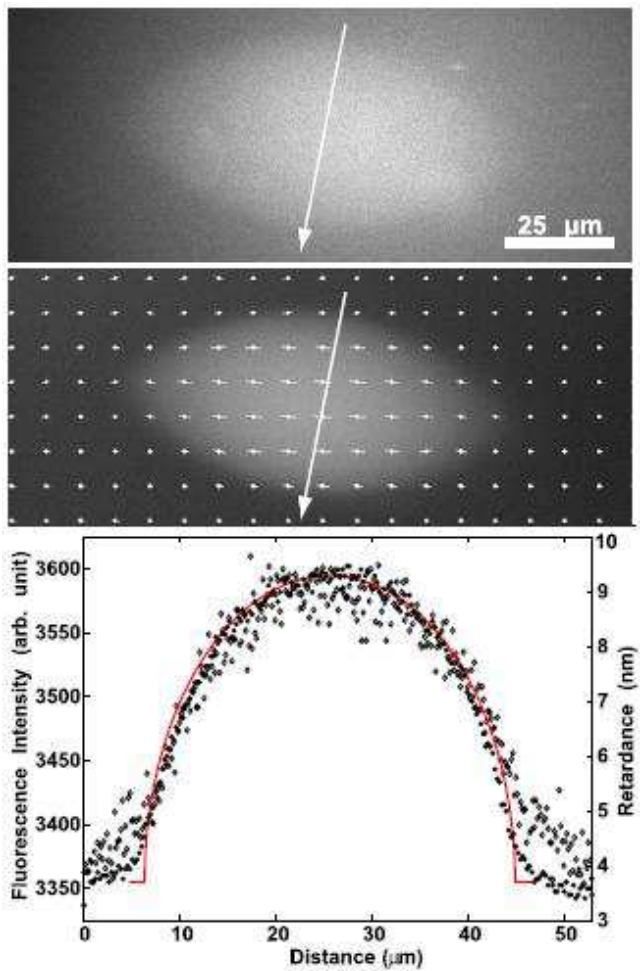}
\caption{\label{fig2}A tactoid measured by fluorescence microscopy (top) and birefringence microscopy (middle) in a solution of 10.8 mg/ml F-actin.  The arrows depict line scans performed across the width of the tactoid and the 25 $\mu$m scale bar applies to both images.  Fluorescence intensity ($\diamond$), birefringence retardance ($\bullet$), and a fit of a scaled semicircle to the birefringence data (solid line) are plotted.  The plot illustrates the agreement between the two methods of imaging, both showing the three dimensional symmetry of the tactoid.  The fluorescence image reveals distinct concentrations in each region and yields a result 36 $\pm$ 5\%  greater inside the tactoid than the surrounding solution.  The birefringence image shows a clear difference in alignment inside the nematic tactoid and the surrounding solution.}
\end{figure}

Quantitative fluorescence microscopy has been performed on selected tactoids (\emph{top}, Fig.~\ref{fig2}).  The fluorescence image clearly shows a difference in concentration between the tactoid and the surrounding solution.  This fluorescence intensity data allows us to quantitatively determine the difference in concentration between the nematic and isotropic regions.  We have measured a number of N tactoids and found that the N region has a concentration 36 $\pm$ 5\% greater than the background I medium \footnote{Coexistence concentrations were determined based on two known properties.  First, the tactoid has rotational symmetry about the long axis.  This property was confirmed through confocal imaging as well as the functional fit in Fig.~\ref{fig2}.  Second, within the thickness of the capillary the labeled protein along the direction perpendicular to the flat face contribute nearly equally to the local fluorescence intensity reading.  This was confirmed by adjusting the focal position and noting negligible changes when the top, middle, or bottom layer of the sample was best in focus.  Thus by knowing the thickness of the capillary, the thickness of the tactoid and the intensity values inside and outside of the tactoid, we obtained the relative protein concentrations in the nematic and isotropic regions.}.  Measurements reveal that I tactoids in a N background show similar concentration differences.  These results confirm the first order nature of the I-N transition. 

Additional properties of the tactoids are obtained via birefringence imaging (\emph{middle}, Fig.~\ref{fig2}).  Average alignment is represented by a line segment through each white dot where the length of the segment depicts the magnitude of the retardance.  The I solution surrounding the tactoid shows no average alignment and clearly illustrates the difference in alignment between the tactoid and its surrounding regions.  Slow axis measurements inside the N tactoids suggest that the director field is bipolar, connecting two point defects, termed boojums \cite{Prinsen_vanderSchoot}, at the opposite poles.  Retardance data plotted below the image, shown with a fit to an equation scaled from a semicircle \footnote{The data is fit to $y=A + B(C^2-(x-D)^2)^{1/2}$.  \emph{A} and \emph{D} shift the center of the circle, \emph{B} is a pre-factor to account for different units, and \emph{C} is the radius of the circle at the middle cross section.}, support the three dimensional characterization of the tactoid.    

\begin{figure}[t]
\includegraphics[width=8.5cm]{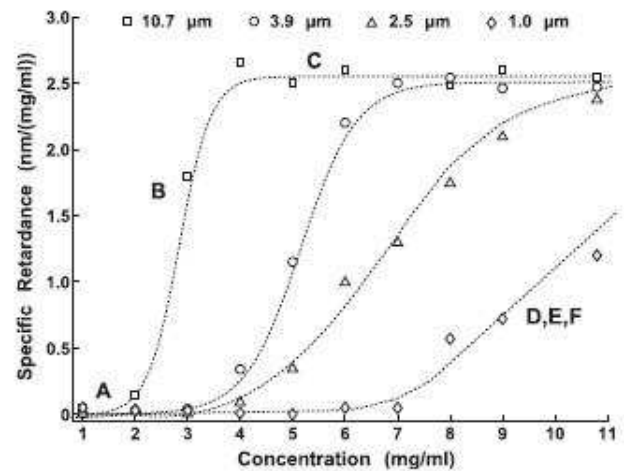}
\caption{\label{fig3}The specific retardance of F-actin is plotted as a function of concentration with varying $\ell$. An F-actin solution with no added gelsolin ($\square$) and with average filament lengths of 3.9$\mu$m ($\circ$), 2.5 $\mu$m ($\vartriangle$), and 1.0 $\mu$m ($\diamond$). The letter designations correspond to regions denoted in Fig.~\ref{fig1}. The dashed lines are guides for the eye.}
\end{figure}

Using this technique we have also measured the average alignment of F-actin at four $\ell$ as a function of actin concentration over the range of the I-N transition (Fig.~\ref{fig3}). The $\ell$ was determined by fluorescence imaging or AFM (for the shortest $\ell$) of single F-actin for at least 500 filaments of each $\ell$. Below a threshold concentration, F-actin solutions are in the I phase, thus producing zero retardance (Region \textbf{A}). As the  concentration increases, the solution reaches the I-N transition region, characterized by a sharp increase of specific retardance (Region \textbf{B}). In the high concentration region, F-actin solution is completely in the N phase (Region \textbf{C}). As $\ell$ decreases, the onset concentration of the I-N transition increases, consistent with the earlier reports \cite{Coppin, Furakawa, Suzuki, Viamontes2003}.  \textbf{D, E,} and \textbf{F} in Fig. \ref{fig3} indicate the I-N transition region for $\ell$ = 1.0 $\mu$m, where the specific birefringence values were measured prior to phase separation as shown in Fig. \ref{fig1}. These birefringence measurements help define the range of concentrations over which the I-N transition occurs.

When actin is polymerized at slightly above the I concentration, nascent tactoids are nucleated in the transition region for F-actin with $\ell \leq$ 2 $\mu$m. Coalescence of tactoidal droplets is often observed, along with the continuous growth of separate tactoids over the time span of hours. When the actin concentration is close to midway between the co-existing I and N concentrations (both are sensitive functions of $\ell $), we reproducibly observed the phenomenon of spinodal decomposition.  The details of this are the subject of further investigation. Briefly, at the high concentration of several mg/ml, F-actin polymerization occurs within seconds following the addition of KCl and MgCl$_2$ \cite{Carlier1984}. The solution, therefore, becomes weakly aligned by the shear flow as it is injected into the capillary tube. A granular structure appears throughout the capillary tube within minutes after the initiation of polymerization. By the 30 minute time point, both I and N droplets are discernable. The late stage growth lasts for many hours, with both continuous growth of existing tactoids and occasional coalescence of neighboring tactoids. 

Several features which suggest metastability are observed in the I-N transition region.  While the shape of the tactoids is consistent throughout the sample, there is appreciable scatter in the data comparing the dimensions of the droplets (not shown). Furthermore, measurements indicate that the local concentration within the I and N regions shows a dependence on the total average concentration.  For a true first order phase transition, the volume fraction of the solution in the I or N phase would depend on the total average concentration of the solution, but the local concentration of each phase would remain constant.  Both I and N tactoids are also regularly found to be present in a weakly aligned background.   taken together these observations lead us to conclude that the system often exists in a metastable state.

It is important to note that the actin tactoids we report here are fundamentally different from what we previously reported \cite{TangHyeran2005}. Actin granules characterized in our previous report were induced by an actin cross-linking protein, alpha-actinin. As a result, the density of actin in the actin granules is over 10 times higher than in this study. The tactoids formed here due to the I-N separation also easily disappear upon dilution and slight agitation, unlike the permanent tactoids due to the cross-linking of alpha-actinin. The similarity in shape likely has to do with a minimization of the free energy density where there is a competition between the splay and bending elastic energies and the surface tension of a droplet \cite{Prinsen_vanderSchoot,Kaznacheev_Bogdanov_Taraskin}. It is thus not surprising that tactoids of similar shapes are found in two distinct types of actin granules, in concentrated suspensions of Tobacco Mosaic Virus (TMV) \cite{TMV}, and in that of filamentous phage fd \cite{DogicPRL2004}.

In conclusion, we have shown new features of the I-N phase transition of F-actin solutions as a function of $\ell$. At $\ell \geq$ 2 $\mu$m, the I-N phase transition is continuous, consistent with previous findings \cite{Viamontes2003,Coppin}. However, biphasic behavior characteristic of a first order transition is observed for $\ell \leq$ 2 $\mu$m, including both phenomena of nucleation-growth and spinodal decomposition. A clear separation of phases with different concentrations is shown for the first time.  Tactoidal droplets of either I or N domains form in the N or I background, respectively. Tactoids of both phases can also coexist with a weakly aligned background state, suggesting metastability. The process for tactoidal growth involves both constant recruitment of the surrounding filaments and coalescence of existing neighboring tactoids.

\acknowledgments
This work is supported by the National Science Foundation (NSF) awards DMR 0405156 and DMR 0079964 (MRSEC), and the Petroleum Research Fund (PRF 42835-AC7), administered by the American Chemical Society. We thank Professors Robert Meyer, Robert Pelcovits, Thomas Powers and James Valles for valuable suggestions.

\end{document}